\documentclass[12pt]{article}
\usepackage{amsfonts}
\usepackage{amsfonts}
\usepackage{amssymb}
\usepackage{amscd}

\date{}

\begin{document}

 \title{\Large{\bf Some $p$-Adic  Aspects of Superanalysis\footnote{\textsf{\,
 Based on the talk presented at the International Workshop {\it Supersymmetries and Quantum Symmetries},
 24-29 July 2003, Dubna, Russia}}  }}

\author{Branko
Dragovich\,\footnote{\textsf{\, E-mail:\,dragovich@phy.bg.ac.yu}}\\
 {\it Institute of Physics, P.O. Box 57,} \\ {\it 11001
Belgrade, Serbia and Montenegro}}

\maketitle

\begin{abstract}
A brief review of a superanalysis over real and $p$-adic
superspaces is presented. Adelic superspace is introduced and an
adelic superanalysis, which contains real and $p$-adic
superanalysis, is initiated.
\end{abstract}

\bigskip

\bigskip

\noindent{\bf 1 Introduction}

\bigskip
Supersymmetry plays very important role in construction of new
fundamental models of high energy physics beyond the Standard
Model. Especially it is significant in formulation of
String/M-theory, which is presently the best candidate for
unification of matter and interactions. Supersymmetry
transformation can be regarded as transformation in a superspace,
which is an ordinary spacetime extended by some anticommuting
(odd) coordinates. Spacetime in M-theory is eleven-dimensional
with the Planck length as the fundamental one. According to the
well-known uncertainty
\begin{equation}
\Delta x \geq \ell_0 = \sqrt {\frac{\hbar G}{c^3}} \approx
10^{-33} cm, \label{1.1}
\end{equation}
one cannot measure distances smaller than the Planck length
$\ell_0$. Since the derivation of (\ref{1.1}) is based on the
general assumption that real numbers and archimedean geometry are
valued at all scales it means that the usual approach is broken
and cannot be extended beyond the Planck scale without adequate
modification which contains non-archimedean geometry. The very
natural modification is to use adelic approach, since it contains
real and $p$-adic numbers which make all possible completions of
the rational numbers. As a result it follows that one has to
consider possible relations between adelic and supersymmetry
structures. In this report we review some aspects of $p$-adic
superanalysis and introduce adelic superanalysis, which is a basis
for investigation of the corresponding $p$-adic and adelic
supersymmetric models.

\bigskip

\bigskip

\noindent{\bf 2 Some basic properties of $p$-adic numbers and
adeles }

\bigskip
Let us first recall that numerical experimental results belong to
the field of rational numbers $\mathbb Q$. On the $\mathbb Q$ one
can introduce the usual absolute value $|\cdot|_\infty$ and
$p$-adic absolute value $|\cdot|_p$ for each prime number $p$.
Completion of $\mathbb Q$ with respect to $|\cdot|_\infty$ gives
the field of real numbers ${\mathbb Q}_\infty \equiv \mathbb R$.
If we replace $|\cdot|_\infty$ by $|\cdot|_p$ then completion of
$\mathbb Q$ yields a new  number field known  as the field of
$p$-adic numbers ${\mathbb Q}_p$. Consequently, $\mathbb Q$ is
dense in $\mathbb R$ as well as in ${\mathbb Q}_p$ for every $p$.
${\mathbb R}$  has archimedean metric $d_\infty (x,y)
=|x-y|_\infty$ and ${\mathbb Q}_p$ has non-archimedean  metric
(ultrametric) $d_p (x,y)=|x-y|_p$ , i.e. $d_p(x,y)\leq max
\{d_p(x,z), d_p(z,y) \}$. It is worth pointing out that ${\mathbb
R}$ and ${\mathbb Q}_p$ exhaust all possibilities to get number
fields by completion of $\mathbb Q$. Any $p$-adic number $x \in
{\mathbb Q}_p $ can be presented in the unique way $x= p^\nu \,
\sum_{k=0}^\infty a_k p^k$ where $\nu \in {\mathbb Z} , \, \, a_k
\in \{0, 1, \cdots, p-1  \}$ , what resembles representation of a
real number $y= \pm\, 10^\mu \, \sum^{k=0}_{-\infty} b_k 10^k ,\,
\, \mu \in {\mathbb Z} , \, \, b_k \in \{0, 1, \cdots, 9  \}$ but
with expansion in the opposite way. There are  two main types of
functions with $p$-adic argument: $p$-adic valued and real-valued
(or complex-valued). The reader who is not familiar with $p$-adic
numbers and their functions can see,  e.g. \cite{vladimirov1}.

To regard simultaneously real and $p$-adic properties of rational
numbers and their completions one uses concept of adeles. An adele
$x$ (see, e.g. \cite{gelfand}) is an infinite sequence $
  x= (x_\infty, x_2, \cdots, x_p, \cdots), $
where $x_\infty \in {\mathbb R}$ and $x_p \in {\mathbb Q}_p$ with
the restriction that for all but a finite set $\bf S$ of primes
$p$ one has  $x_p \in {\mathbb Z}_p = \{x \in {\mathbb Q}_p\, : |x
|_p \leq 1 \}= \{x \in {\mathbb Q}_p\, : x = a_0 + a_1 p + a_2 p^2
+ \cdots \} $. Componentwise addition and multiplication endow a
ring structure to the set of adeles ${\cal A}$. ${\cal A}$ can be
defined as
\begin{equation}
 {\cal A} = \bigcup_{{\bf S}} {\cal A}_{\bf S},
 \ \  {\cal A}_{\bf S} = {\mathbb R}\times \prod_{p\in {\bf S}} {\mathbb Q}_p
 \times \prod_{p\not\in {\bf S}} {\mathbb Z}_p \, .         \label{1.2}
\end{equation}
$\mathbb Q$ is naturally embedded in ${\cal A}$. Ring ${\cal A}$
is also a locally compact topological space. Important functions
on ${\cal A}$ are related to mappings $f: {\cal A}\to {\cal A} $
and $\varphi: {\cal A} \to {\mathbb R}\, (\mathbb C)$.

\bigskip

\bigskip

\noindent{\bf 3 Elements of $p$-adic  and adelic string theory }

\bigskip
A notion of $p$-adic string and  hypothesis on the existence of
non-archimedean geometry at the Planck scale were introduced by
Volovich \cite{volovich1} and have been investigated by many
researchers (reviews of an early period are in \cite{vladimirov1}
and \cite{brekke}). Very successful $p$-adic analogues of the
Veneziano and Virasoro-Shapiro amplitudes were proposed in
\cite{freund1} as the corresponding Gel'fand-Graev \cite{gelfand}
beta functions. Using this approach, Freund and Witten obtained
\cite{freund2} an attractive adelic formula $
 A_\infty (a,b) \prod_p A_p (a,b) =1 ,
$ which states that the product of the crossing symmetric
Veneziano (or Virasoro-Shapiro) amplitude and its all $p$-adic
counterparts equals unity (or a definite constant). This gives
possibility to consider an ordinary four-point function, which is
rather complicate, as an infinite product of its inverse $p$-adic
analogues, which have simpler forms. The ordinary crossing
symmetric Veneziano amplitude can be defined by a few equivalent
ways and its integral form is
\begin{equation}
  A_\infty(a,b)        =  \int_{{\mathbb R}} \vert x \vert_\infty^{a-1}
  \, \vert 1-x\vert_\infty^{b-1} \, dx ,
     \label{1.3}
\end{equation}
where it is taken $\hbar=1,\ T=1/\pi$, and $a=-\alpha (s) = - 1
-\frac{s}{2}, \ b=-\alpha (t), \ c=-\alpha (u)$ with the
conditions $s+t+u = -8$ and $a+b+c=1$. According to \cite{freund1}
$p$-adic Veneziano amplitude is a simple $p$-adic counterpart of
(\ref{1.3}), i.e.
\begin{equation}
  A_p(a,b)        =  \int_{{\mathbb Q}_p} \vert x \vert_p^{a-1}
  \, \vert 1-x\vert_p^{b-1}\,  dx ,
     \label{1.4}
\end{equation}
where now $x \in  {\mathbb Q}_p$. In both (\ref{1.3}) and
(\ref{1.4}) kinematical variables $a, b, c$ are real or
complex-valued parameters. Thus in (\ref{1.4}) only string
world-sheet parameter $x$ is treated as $p$-adic variable, and all
other quantities maintain their usual  real values. Unfortunately,
there is a problem to extend the above product formula to the
higher-point functions. Some possibilities to construct $p$-adic
superstring amplitudes are considered in \cite{arefeva1} (see also
\cite{brekke1}, \cite{ruelle1},  and \cite{vladimirov2}).

 A recent interest in $p$-adic string theory has been mainly
related to an extension of adelic quantum mechanics
\cite{dragovich} and $p$-adic path integrals to string amplitudes
\cite{dragovich1}. An effective nonlinear $p$-adic string theory
(see, e.g. \cite{brekke}) with an infinite number of space and
time derivatives has been recently of a great interest in the
context of the tachyon condensation \cite{sen}. It is also worth
mentioning successful formulation and development of $p$-adic and
adelic quantum cosmology (see \cite{dragovich2} and references
therein) which demonstrate discreteness of minisuperspace with the
Planck length $\ell_0$ as the elementary one.

\bigskip

\bigskip

\noindent{\bf 4 Elements of $p$-adic and adelic superanalysis }

\bigskip
Here it will be first presented  some elements of real and
$p$-adic superanalysis along approach introduced by Vladimirov and
Volovich \cite{vladimirov3} and elaborated  by Khrennikov
\cite{khrennikov}. Then I shall generalize this approach to adelic
superanalysis.

Let $\Lambda ({\mathbb Q}_v ) = \Lambda_{0} ({\mathbb Q}_v)\,
\oplus \, \Lambda_{1} ({\mathbb Q}_v)$ be $Z_2$-graded vector
space over ${\mathbb Q}_v , \, \, (v = \infty, 2, 3, \cdots, p,
\cdots)$, where elements  $a \in\Lambda_{0} ({\mathbb Q}_v)$ and
$b\in \Lambda_{1} ({\mathbb Q}_v)$ have even $(p(a) = 0)$ and odd
$(p(b)=1)$ parities. Such $\Lambda ({\mathbb Q}_v)$ space is
called $v$-adic (real and $p$-adic) superalgebra if it is endowed
by an associative algebra with unity and parity multiplication
$p(ab) \equiv p(a) + p(b) \, (mod\, 2)$. Supercommutator is
defined in the usual way: $[ a , b \} = a\, b - (-1)^{p(a) p(b)}
b\, a$. Superalgebra $\Lambda ({\mathbb Q}_v)$ is called
(super)commutative if $[ a , b \} =0$ for any $a\in \Lambda_{0}
({\mathbb Q}_v)$ and $b \in \Lambda_{1} ({\mathbb Q}_v)$. As
illustrative examples of commutative superalgebras one can
consider finite dimensional $v$-adic Grassmann algebras $G (
{\mathbb Q}_v : \eta_1, \eta_2, \cdots, \eta_m)$ which dimension
is $2^m$ and generators $\eta_1, \eta_2, \cdots, \eta_m$ satisfy
anticommutative relations $ \eta_i \eta_j + \eta_j \eta_i = 0$.
The role of norm necessary to build analysis on commutative
superalgebra $\Lambda ({\mathbb Q}_v)$ plays the absolute value
$|\cdot|_\infty$ for real case and $p$-adic norm $|\cdot|_p$ for
$p$-adic cases.

Let $\Lambda ({\mathbb Q}_v)$ be a fixed commutative $v$-adic
superalgebra. $v$-Adic superspace of dimension $(n,m)$ over
$\Lambda ({\mathbb Q}_v)$ is
\begin{equation}  {\mathbb Q}_{\Lambda
({\mathbb Q}_v)}^{n,m} = \Lambda_{0}^n ({\mathbb Q}_v) \, \times
\Lambda_{1}^m ({\mathbb Q}_v) \label{4.1} \end{equation} and it is
an extension of the standard $v$-adic space. In the sequel we will
mainly have in mind that $\Lambda_{0} ({\mathbb Q}_v)= {\mathbb
Q}_v$ or that  ${\mathbb Q}_v$ is replaced by  ${\mathbb Q}_v
(\sqrt{\tau})$, where $ \sqrt{\tau} \not\in {\mathbb Q}_v .$  Then
our $v$-adic (i.e. real and $p$-adic) superspace can be defined as
${\mathbb Q}_{\Lambda ({\mathbb Q}_v)}^{n,m} = {\mathbb Q}_v^n \,
\times \Lambda_{1}^m ({\mathbb Q}_v)$ which points are $X^{(v)} =
(X^{(v)}_1, X^{(v)}_2,\cdots, X^{(v)}_n, X^{(v)}_{n+1}, \cdots,
X^{(v)}_{n+m}) = (x^{(v)}_1, x^{(v)}_2, \cdots, x^{(v)}_n,
\theta^{(v)}_1, \cdots, \theta^{(v)}_m)$ $ = (x^{(v)},
\theta^{(v)})$, where coordinates $x^{(v)}_1, x^{(v)}_2, \cdots,
x^{(v)}_n$ are commutative, $p(x^{(v)}_i) =0$, and
$\theta^{(v)}_1, \theta^{(v)}_2 \cdots, \theta^{(v)}_m$ are
anticommutative (Grassmann), $p(\theta^{(v)}_j) = 1$. Since
supercommutator $\,[ X^{(v)}_i , X^{(v)}_j \} = X^{(v)}_i\,
X^{(v)}_j -\,\break (-1)^{p(X^{(v)}_i) p(X^{(v)}_j)}\, X^{(v)}_j
\, X^{(v)}_i = 0 ,$ coordinates $X^{(v)}_i , \, \, (i = 1, 2,
\cdots, n+m )$ are called supercommuting. A norm of $X^{(v)}$ can
be defined as $||X^{(v)}|| = max\{ \, |x^{(v)}_i|_v  \, ,
|\theta^{(v)}_j |_v \, \}$. In the sequel, to decrease number of
indices we often omit them when they are understood from the
context.

One can define functions $F_v (X)$  on open subsets of superspace
${\mathbb Q}_{\Lambda ({\mathbb Q}_v)}^{n,m} $, as well as their
continuity and differentiability (for some details, see
\cite{vladimirov3} and \cite{khrennikov}). One has to differ the
left and the right partial derivatives: $\frac{\partial_L
F_v}{\partial \theta_j}\, , \, \frac{\partial_R F_v}{\partial
\theta_j}$. It is worth noting that derivatives of $p$-adic valued
function of $p$-adic arguments are formally the same as those for
real functions of real arguments. Integral calculus for $p$-adic
valued functions is more subtle than in the real case, since there
is no $p$-adic valued Lebesgue measure \cite{schikhof}. One can
use antiderivatives, but one has to take care about
pseudoconstants, which are some exotic functions with zero
derivatives. However, for analytic functions one can well define
definite integrals using the corresponding antiderivatives
\cite{vladimirov1}. Integration with anticommuting variables  is
introduced by axiomatic approach requiring linearity and
translation invariance in both real and $p$-adic cases. In
particular, one obtains the following two indefinite integrals:
$\int d\, \theta^{(v)}_j = 0$ and $\int \theta^{(v)}_j \, d\,
\theta^{(v)}_j = 1$.

When ${\mathbb Q}_v^n$ corresponds to an $n$-dimensional
spacetime,  functions $F_v (x, \theta)$ on superspace ${\mathbb
Q}_{\Lambda ({\mathbb Q}_v)}^{n,m}$ are called  $v$-adic
superfields. Due to the fact that there is only finite number of
non-zero products with anticommuting variables, expansions of $F_v
(x, \theta)$ over $\theta_j,  \, \, (j=1,2, \cdots , m) $ are
finite, i.e. there are $2^m$ terms in the corresponding Taylor
expansion. Description of supersymmetric models by superfields is
very compact and elegant \cite{wess}.

We can now turn to adelic superanalysis. It is natural to define
the corresponding $Z_2$-graded  vector space over ${\cal A}$ as
\begin{equation}
\Lambda ({\cal A}) = \bigcup_{\bf S} \Lambda_{\bf S}  \, , \,
\quad \Lambda_{\bf S}  = \Lambda ({\mathbb R}) \times \prod_{p \in
{\bf S}} \Lambda ({\mathbb Q}_p) \times \prod_{p \not\in {\bf S}}
\Lambda ({\mathbb Z}_p ) \, ,\label{4.2}
\end{equation}
where $\Lambda ({\mathbb Z}_p )  = \Lambda_0 ({\mathbb Z}_p
)\oplus \Lambda_1 ({\mathbb Z}_p )$ is a graded vector space over
the ring of $p$-adic integers $ {\mathbb Z}_p  $ and ${\bf S}$ is
a finite set of primes $p$. Graded vector space (\ref{4.2})
becomes adelic superalgebra by requiring that $ \Lambda ({\mathbb
R}) \, , \Lambda ({\mathbb Q }_p ) \, , \Lambda ({\mathbb Z}_p )$
are superalgebras. Adelic supercommutator may be regarded as a
collection of real and all $p$-adic supercommutators. Thus adelic
superalgebra (\ref{4.2}) is commutative.  An example of
commutative adelic superalgebra is the following adelic Grassmann
algebra:
$$ G ({\cal A}: \eta_1, \eta_2, \cdots, \eta_m) = \bigcup_{\bf S}
G_{\bf S} ( \eta_1, \eta_2, \cdots, \eta_m)\, $$
\begin{equation}
G_{\bf S} (\eta_1,  \cdots, \eta_m) = G ({\mathbb R} : \eta_1,
\cdots, \eta_m) \times  \prod_{p \in{\bf S}} G ({\mathbb Q}_p
\break : \eta_1,  \cdots, \eta_m)  \times \prod_{p \not\in{\bf S}}
G( {\mathbb Z}_p :\eta_1,  \cdots, \eta_m) . \label{4.3}
\end{equation}

Adelic superspace of dimension $(n,m)$ has the form
\begin{equation}
{\cal A}^{n,m}_{\Lambda ({\cal A})} = \bigcup_{\bf S} {\cal
A}^{n,m}_{\Lambda ({\cal A}),\bf S}\, , \quad  {\cal
A}^{n,m}_{\Lambda ({\cal A}),\bf S} = {\mathbb R}_{\Lambda
({\mathbb R})}^{n,m} \times \prod_{p\in {\bf S}} {\mathbb
Q}_{\Lambda ({\mathbb Q}_p)}^{n,m}  \times \prod_{p\not\in {\bf
S}} {\mathbb Z}_{\Lambda ({\mathbb Z}_p)}^{n,m}\, , \label{4.4}
\end{equation}
where ${\mathbb Z}_{\Lambda ({\mathbb Z}_p)}^{n,m} $ is
$(n,m)$-dimensional $p$-adic superspace over superalgebra
${\Lambda ({\mathbb Z}_p)}$. Closer to supersymmetric models is
the  superspace ${\cal A}^{n,m}_{\Lambda ({\cal A})} =
\bigcup_{\bf S} {\cal A}^{n,m}_{\Lambda ({\cal A}),\bf S}\,$ where
\begin{equation}
{\cal A}^{n,m}_{\Lambda ({\cal A}),\bf S} = ({\mathbb R}^n \times
{\Lambda_1^m ({\mathbb R})}) \times \prod_{p\in {\bf S}} ({\mathbb
Q}_p^n \times {\Lambda_1^m ({\mathbb Q}_p)})  \times
\prod_{p\not\in {\bf S}} ({\mathbb Z}_p^n \times {\Lambda_1^m
({\mathbb Z}_p)}) . \label{4.5}
\end{equation}
Points of adelic superspace $X$ have the  coordinate form $X =
(X^{(\infty)}, X^{(2)}, \cdots ,\break X^{(p)}, \cdots )$, where
for all but a finite set of primes ${\bf S}$ it has to be
$||X^{(p)}|| = max\, |X^{(p)}_i|_p \, \, \leq 1 .$ The
corresponding adelic valued functions (superfields) must  satisfy
adelic structure, i.e. $F (X) = (F_\infty , F_2, \cdots , F_p ,
\cdots  )$ with condition $|F_p|_p \, \leq 1$ for all but a finite
set of primes ${\bf S}$. In the spirit of this approach one can
continue to build adelic superanalysis.

\bigskip

\bigskip

\noindent{\bf 5 Concluding remarks }

\bigskip
In this report are presented some elements of $p$-adic and adelic
generalization of superanalysis over real numbers. We have been
restricted to superanalysis over the field of $p$-adic numbers
${\mathbb Q}_p$ and the corresponding ring of adeles ${\cal A}$.
It is worth noting that algebraic extensions of ${\mathbb Q}_p$
give much more possibilities than in the real case, where there is
only one extension, i.e. the field of complex numbers ${\mathbb C}
= {\mathbb R}(\sqrt {-1})$. In fact there is at least one  $\tau
\in {\mathbb Q}_p $ such that ${\mathbb Q}_p (\tau^{\frac{1}{n}})
\neq {\mathbb Q}_p$ for a fixed $p$ and any integer $n\geq 2 $.
Note that there are three  and seven $p$-adic distinct quadratic
extensions ${\mathbb Q}_p (\sqrt{\tau})$ if $p \neq 2$ and $p=2$,
respectively. Using these quadratic extensions,  supersymmetric
quantum mechanics with $p$-adic valued functions is constructed by
Khrennikov \cite{khrennikov}. The above approach to adelic
superanalysis may be easily generalized to the case when $
{\mathbb R} \to {\mathbb C} , \, \, {\mathbb Q}_p  \to {\mathbb
Q}_p (\sqrt{\tau}) , \, \, {\mathbb Z}_p  \to {\mathbb Z}_p
(\sqrt{\tau}) $. Algebraically closed and ultrametrically complete
analogue of ${\mathbb C} $ is ${\mathbb C}_p $, which is an
infinite dimensional vector space. Thus $p$-adic algebraic
extensions offer enormously rich  and very challenging field of
research in analysis as well as in superanalysis.

It  is also very desirable to find  a formulation of superanalysis
which would be a basis for supersymmetric generalization of
complex-valued $p$-adic and adelic quantum mechanics
\cite{dragovich} as well as of related  quantum field theory and
Superstring/M-theory \cite{dragovich1}.

\bigskip

\bigskip

\noindent {\bf Acknowledgements\,} The work on this paper was
supported in part by the Serbian Ministry of Science, Technologies
and Development under contract No 1426 and by RFFI grant
02-01-01084 .

\end{document}